\documentstyle[preprint,revtex]{aps}

\def\phys{Phys. Rev. B}
\def\lett{Phys. Rev. Lett.}

\begin{document}
\draft

\begin{title}
{\bf Flat quasiparticle dispersion \\
and a hidden small energy scale \\
in the 2D t-J model} \\
\end{title}

\author{Elbio~Dagotto$^1$, Alexander~Nazarenko$^1$
and Massimo~Boninsegni$^2$ }

\begin{instit}
$^1$Department of Physics and
National High Magnetic Field Laboratory,      \\
Florida State University,
Tallahassee, FL 32306. \\

$^2$ National Center for Supercomputing Applications, \\
University of Illinois at Urbana-Champaign, \\
405 N. Mathews Avenue, Urbana, IL 61801. \\
\end{instit}

\receipt{February, 1994}

\begin{abstract}
A model of weakly interacting hole quasiparticles is proposed to
describe the
normal state of the high temperature
superconductors. The effect of strong correlations is contained in the
dispersion relation of the holes, which is obtained accurately using
a numerical technique and the t-J model on $12\times 12$ clusters.
Saddle-points generated by many-body effects
induce quasiparticle
$flat$ bands similar
to those observed in recent angle-resolved photoemission experiments.
The near degeneracy between momentum states along
the ${\rm X-Y}$ direction is a hidden small energy scale
in the problem, which produces a
strongly temperature dependent Hall coefficient, as well as
a large Fermi surface similar to those observed in the cuprates.

\end{abstract}



\newpage

Many normal state properties of the high temperature superconductors
are still puzzling.
The in-plane d.c. resistivity, $\rho_{ab}$,
of the hole-doped cuprates
is linear with
temperature (T) when the hole concentration is optimal i.e. when
the critical temperature (T$_c$) is the highest.\cite{batlogg}
The Hall coefficient, ${\rm R_H}$, is strongly temperature dependent
in the range between
T$_c$ to $T \sim 800K$. At
constant temperature, ${\rm R_H}$ changes sign as the hole  density is
increased away from the insulator parent compound.\cite{ong,hwang}
Angle-resolved photoemission experiments (ARPES) suggest the
presence of a large (electron-like) Fermi surface in some
cuprates.\cite{review}
This result is in contradiction with
theoretical ideas at T=0 based on electronic models
which tend to favor the presence of small hole-like Fermi
surfaces at low hole density.\cite{bob}
In addition,
using ARPES techniques it has been recently reported\cite{dessau} that
an extended region of flat ${\rm Cu
O_2}$-derived bands very near the Fermi energy exist for Bi2212,
Bi2201, Y123 and Y124. This unexpected result cannot be explained within band
structure calculations.\cite{dessau}

In this paper we discuss a model of hole carriers in an
antiferromagnetic background that explains in a natural way many
of the above mentioned anomalous properties of the cuprates. The
main assumption is that the normal state regime of
the cuprates can be modeled by a non-interacting gas of
hole-quasiparticles. These carriers are strongly dressed
by spin-waves.
The influence of antiferromagnetism and strong correlations
are contained in the special dispersion relation, $\epsilon ( {\bf k} )$,
which is obtained using an unbiased
numerical method applied to a one band model description of the
${\rm Cu O_2}$ planes.
Our approach is similar in spirit to that of
Trugman\cite{trugman}
where the dispersion relation of holes was calculated variationally
using the ${\rm t t' J}$ model.
One of the main results of the paper is that
the hole dispersion relation contains $flat$ bands at the X and Y points
(i.e. ${\bf k} = (\pi,0)$, and
$(0, \pi)$), very similar to those described in the recent ARPES
experiments.\cite{dessau}
The present approach
belongs to the family of ``narrow band'' descriptions of the normal
state of the cuprates, with flat bands caused by
strong electronic correlations instead of band structure effects.
In addition, $\epsilon ( {\bf k} )$
contains a
somewhat hidden energy scale, $\Delta$, which is considerably smaller than
the total quasiparticle bandwidth of the order
of the antiferromagnetic exchange ${\rm J}$.
While this small
scale was observed in previous calculations by
Trugman,\cite{trugman}, Dagotto et al.,\cite{viejo}
and others, its physical
consequences have not been compared with recent experiments for
the cuprates. We show that the strong temperature dependence of ${\rm
R_H}$ and the Fermi surface (FS) induced by the presence
of $\Delta$ produces results in good agreement with the data
obtained by Hwang et al.\cite{hwang}

To calculate
$\epsilon ( {\bf k} )$, here we used the two dimensional ${\rm t-J}$
model.\cite{rice}
However, the intuitive ideas to be
discussed below are not dependent on the specific model selected
to represent the cuprates, as long as strong
antiferromagnetic fluctuations are present in the ground state.
Using the one band Hubbard model near
half-filling, the results of our analysis are qualitatively the same.
The ${\rm t-J}$  model Hamiltonian is defined as
\begin{eqnarray}
{\rm H = -t \sum_{ \langle {\bf i}{\bf j} \rangle }
[ ( 1 - n_{{\bf i}, -\sigma} )
{c}^{\dagger}_{{\bf i}\sigma}
{c}_{{\bf j}\sigma} ( 1 - n_{{\bf j}, -\sigma} )  + h.c. ]
 + J \sum_{ \langle {\bf i}{\bf j} \rangle }
[ {\bf S}_{{\bf i}}.{\bf S}_{{\bf j}} - {{1}\over{4}} n_{\bf i} n_{\bf j} ] },
\end{eqnarray}
\noindent
where the notation is standard.
The energy of
holes in this model has been extensively analyzed in the literature
using a wide variety of numerical
and analytical techniques.\cite{review}
Here $\epsilon ( {\bf k} )$ is calculated using a
Green Function Monte Carlo (GFMC) method\cite{massimo} which allows to
obtain results on large clusters of $12\times 12$ sites minimizing finite
size effects.
We have explored other techniques to carry out this
calculation (Lanczos method on clusters from 16 to 26 sites, and the Born
approximation at large spin S),
and all of them produce results in good agreement with GFMC.


In Fig.1, the numerically evaluated  $\epsilon ( {\bf k} )$
is shown for ${\rm J/t=0.4}$ along particular directions in the
Brillouin zone.
The total bandwidth, ${\rm W}$, is severely reduced from that of a gas
of non-interacting electrons
due to the antiferromagnetic correlations. The
minimum in the energy is obtained at the ${\bar M}$ point
${\bf k} = (\pi/2, \pi/2)$.
These observations are in agreement with
a wide variety of
previous calculations,\cite{trugman,review} which have consistently obtained a
bandwidth
${\rm W \sim 2J}$ and the energy minimum at ${\bar M}$.
However, the analysis below suggests that the strong $anisotropy$ of
$\epsilon ( {\bf k} )$ is more important than the actual
bandwidth for calculations carried out near room temperature.
The close proximity in energy of momenta
${\bar M}$ and X
introduces a small energy scale $\Delta = \epsilon(X) - \epsilon({\bar M})$ in
the
problem in agreement with previous variational and numerical
calculations.\cite{trugman,viejo}
Actually, all momenta
belonging to the non-interacting 2D Fermi surface
${\rm cosk_x + cosk_y = 0}$ are
very close in energy in the ${\rm t-J}$ model.
In the range ${\rm 0.3 \leq J/t \leq 0.7}$,
$\Delta$ is approximately
15\% to 20\% of the total bandwidth.
Since ${\rm W \sim J}$, then important effects
are expected at
$T \sim 300K$ i.e. room temperature.


In order to calculate some observables,
we have fit the numerical results of Fig.1 using a general combination
of trigonometric functions. The best fit corresponds to
\begin{eqnarray}
{\rm \epsilon( {\bf k} ) = -1.255 + 0.34~cosk_x~cosk_y +
0.13 (cos2k_x + cos2k_y ) },
\end{eqnarray}
\noindent where higher order harmonics carry negligible weight.
The main $effective$ contribution to $\epsilon ( {\bf k})$ arises from
hole hopping between sites belonging to
the same sublattice, to avoid distorting the AF background.
For completeness, we have also calculated numerically
$\epsilon ( {\bf k})$ when a n.n.n. hopping term of strength ${\rm t'}$
along the diagonals
of the square lattice is added to Eq.(1).
Using ${\rm t'/t = -0.15}$,\cite{hyber}
the energy minimum is now located at X as
shown in Fig.1, and the mass anisotropy between directions
along and perpendicular to
X-Y, is even higher than
at ${\rm t'=0}$.\cite{effect,arno}
However, the actual position of the hole minimum
is not crucial for our arguments below, but more important is the near energy
degeneracy along the ${\rm cosk_x + cosk_y = 0}$  line.
Thus, our conclusions are not
affected by a nonzero ${\rm t'}$.


An important detail of Fig.1 is the near flatness of the energy
in the vicinity
of ${\bf k} = (\pi,0)$. This feature is in striking
agreement with ARPES results obtained by Dessau et al.\cite{dessau} for
Bi2212 (reproduced in Fig.2a).
These authors remarked that such a $flat$ region near
$(0,\pi), (\pi,0)$ is observed in several high-Tc compounds
and seems a universal property of the hole-doped cuprates. To compare the
theoretically observed flat region of Fig.1 and the experiments,
we fixed the chemical potential of our model such that ${\bf k} = (\pi,0)$
would approximately be the Fermi momentum ${\rm p_F}$ and to set the
overall scale we used ${\rm t = 0.4 eV}$.
Only momenta from $\Gamma$ to the Fermi
momentum are plotted since for momenta above ${\rm p_F}$, the data points of
Fig.1 represent poles in the hole spectral function that do not necessarily
carry a large weight, and thus presumably they do not correspond to the
large quasiparticle peaks observed in the ARPES experiment. This issue has been
discussed at
length in previous publications.\cite{dago9292}
The qualitative agreement between the ${\rm t-J}$ model prediction and
experiments
is remarkable showing that the
flatness of the ${\bf k } = (\pi,0)$
region in the cuprates may have a many-body origin.
In the present calculation the presence
of saddle-points near X and Y (also noticed in Ref.\cite{trugman})
seems responsible for the abnormal flatness
of the hole dispersion. The position of the saddle points (SP) at ${\rm J/t
= 0.4}$ is approximately at
${\bf k} = ( 0.73 \pi,0)$ and their rotated ones. From Eq.(2), it can be
shown that the location of the SP near the
X point is a consequence of having the hole-band minimum
at the ${\bar M}$ point. As expected, the density of states,
DOS, presents a van-Hove singularity caused by the saddle-points
(Fig.2b)\cite{trugman}. The
similarities between the present approach, and the van-Hove scenario
widely discussed before in the literature
are remarkable.\cite{newns} The main difference is that in the present
calculation
the generation of the saddle-point in the dispersion is through
$many-body$ effects. The consequences of this saddle-point in the d.c.
resistivity and superconductivity will be discussed in a future publication.

The presence of the small scale $\Delta$ produces
interesting physical consequences. First, let us consider its influence on
Fermi
surface calculations. In Fig.3a, the hole
occupation number in momentum space,
${\rm n_k}$, is shown at 10\% hole density, at several
temperatures ranging from ${\rm T=0 K}$ to ${\rm T \sim 500 K}$.
It is clear that
${\rm n_k}$ at ${\bf k} \approx
(\pi/2, \pi/2)$ rapidly decreases
as the temperature increases. Reciprocally, the
occupation
near ${\bf k} = (\pi,0)$ and $(0,\pi)$ increases. Intuitively, this
result is understandable since different regions in momentum space will start
contributing appreciably to observables once the temperature is
comparable
to their energy with respect to the T=0 chemical potential.
Thus, in our model the near degeneracy between ${\bar M}$ and X implies
that a strong temperature dependence in many quantities should be
expected. In Fig.3b,c the Fermi
surface is shown as a function of temperature. At T=0 and a realistic
hole density ${\rm x = 0.10}$, the minimum of the hole band located at
${\bar M}$ implies the presence of hole-pockets in the Fermi
surface (Fig.3b). However, when the temperature
becomes comparable to $\Delta$ all levels along the X-Y direction
become equally
populated and the hole-pockets are washed out.
To characterize the Fermi surface at a fixed finite temperature,
we decided to plot the lines in k-space where the Fermi distribution
$f$ is appreciably reduced from its maximum value, $f_{max}$,
which is always
obtained at ${\bar M}$. To fix ideas, in Fig.3c we plotted the line
where the population of a given state has been reduced to only
10\% of the maximum $f_{max}$.
At a temperature of the order or
larger than $\Delta$ no vestige of the
hole pockets remain, and the ``Fermi-surface'' is large and
electron-like, as observed experimentally and discussed in the
variational
approach of Trugman.\cite{trugman}



We have also
investigated the effect of a small $\Delta$ on the one band Hubbard model.
While at large ${\rm
U/t}$ the results are quantitatively similar to the t-J model, at small
or intermediate couplings the presence of $\Delta$ has
more dramatic consequences.
For example, at ${\rm U/t=4}$, $\Delta \sim
0.009 t \sim 0.004 eV \sim 42 K $, i.e. approximately an order of magnitude
smaller
than
for the ${\rm t-J}$ model at ${\rm J/t=0.4}$.\cite{arno} Thus, the
disappearance of the hole-pockets near half-filling will occur at an
even lower temperature than for the ${\rm t-J}$ model.
Although the Hubbard model at an intermediate coupling ${\rm
U/t=4}$ may not be a good description of the high-Tc cuprates, our approach
can explain interesting Monte Carlo numerical results
reported in the literature. Working on a ${\rm 16 \times 16}$
cluster, ${\rm T \sim 0.17 t}$ and electronic density $\langle n \rangle =
0.87$, Moreo et al. have observed a large Fermi surface in a numerical
simulation of the Hubbard model.\cite{moreo}
However, our analysis shows that only at temperatures as low as
${\rm T = 0.009t}$
the near degeneracy between ${\bar M}$ and X can be resolved and thus
for all
practical purposes Monte Carlo simulations at currently accessible
temperatures treat all the momenta along X-Y as
degenerate
in energy. Then, the absence of ``hole-pockets'' in the numerical
analysis is now understandable,\cite{horsch} although other possible
many holes
effects that lead to a large Fermi surface can also contribute to the Monte
Carlo simulation results.

The value of ${\rm n_k}$
shown in Fig.3a have important consequences for other
experimentally measurable quantities.
The change from a pocket-like to a large Fermi surface as the
temperature is increased (at low hole density) suggests that ${\rm R_H}$ may
also
acquire a strong temperature dependence in the present model. To check this
idea, ${\rm R_H}$ was calculated in the relaxation-time approximation
as\cite{ziman}
\begin{eqnarray}
{\rm {{R_H}\over{N}} = {{V_0}\over{2eh}}
{{
\sum_{\bf k} (v_x {{\partial v_x}\over{\partial k_y}} - v_y
{{\partial v_y}\over{\partial k_x}} ) (-{{\partial f}\over{\partial E}} )
} \over{
[ \sum_{\bf k} v^2_x ( - {{\partial f}\over{\partial E}} ) ]^2
} } }
\end{eqnarray}
\noindent where N is the number of sites,
${\rm V_0}$ is the volume of the unit cell, and the rest of the notation
is standard. To avoid the complexities associated with
interplane coupling between adjacent ${\rm Cu O_2}$ planes, here we
compare results exclusively with ${\rm La_{2-x} Sr_x Cu O_4}$
at different Sr concentrations.
In this compound, ${\rm V_0 = 3.8 \times 3.8
\times 6.6 \times \AA^3 \approx 95 \times (10^{-8} cm )^3}$.
In Fig.3d, the sign of the Hall coefficient is shown in the plane
T-x. A sign-change occurs at low temperature as
a function of ${\rm x}$ at a concentration ${\rm x \sim 0.4}$, and
near half-filling increasing the temperature at ${\rm T \sim
400K}$.\cite{comm77}
Results for other values of ${\rm J/t}$ in the physically relevant
regime are very similar. In Fig.3d the region where the Fermi surface
changes from hole to electron-like is shown. It is interesting to
observe that there is an intermediate
regime where ${\rm R_H}$ is positive while the FS is
large, which is reminiscent of the experimentally observed behavior in
the cuprates.
Fig.4a shows the hole density dependence
of ${\rm R_H}$ at low temperature compared against recent
experiments by Hwang et al.\cite{hwang}
The agreement theory-experiment is excellent.
Note that on the theoretical side,
no other adjustable parameter is used besides ${\rm J/t}$, and
actually even the ${\rm J/t}$
dependence of the results is weak.
Our simple approach
correctly reproduces the order of magnitude and qualitative trends recently
observed experimentally
in LSCO.
In Fig.4b, the temperature dependence
of ${\rm R_H}$ is shown. At low density, the agreement theory-experiment
is again very good. The shift of the experimental results may be caused
in part by the use of polycrystals (single crystal measurements
systematically
present a smaller ${\rm R_H}$\cite{hwang}). More difficult to understand is
the
behavior at ${\rm x = 0.22}$. While the overall scale of theory and experiment
are similar, the qualitative trends are not. In our model, ${\rm R_H}$
changes sign not only as a function of hole density but also of temperature,
while experimentally ${\rm R_H}$ decays slowly to zero as T is
increased. Quite likely higher energy states not considered in our
model are important to account for this experimental feature.
Nevertheless,
the overall agreement theory-experiment is remarkable.


Summarizing, in this paper we presented a simple model of hole
quasiparticles that accounts for several unusual properties of the
normal state of the cuprates. The effect of strong correlations
is contained in the dispersion relation of holes which is
accurately evaluated using numerical techniques. Results are contrasted
with previous calculations by Trugman where a variational method was
used to calculate the hole dispersion relation. An overall nice
agreement between the two approaches is found. A saddle-point
appears in  $\epsilon ( {\bf k})$, inducing a van-Hove singularity
in the DOS of many-body origin. One of our main results is that
this saddle-point produces a nearly
$flat$ hole band near the X point in agreement with recent ARPES
experiments by Dessau et al.\cite{dessau}
The near degeneracy between X and ${\bar M}$ momenta introduces a
small energy scale in the problem, causing
the Hall coefficient and Fermi surface to have a strong temperature
dependence near room temperature compatible with recent experiments by
Hwang et al.\cite{hwang}.


We have recently received a paper by Bulut et al.\cite{bulut} where
similar discussions for the quasiparticle dispersion are given. We thank
these authors for sending us an advance copy of their paper.
Conversations with D. J. Scalapino,
A. Moreo, Z.-X. Shen, and R. Laughlin
are also gratefully acknowledged.
E. D. is supported by the Office of Naval Research under
grant ONR N00014-93-0495. M. B. is supported by the Office of Naval
Research under grant ONR N00014-J-92-1320.
Numerical calculations were performed in part on a CM-5 supercomputer
at NCSA, Univ. of Illinois, Urbana,
Illinois (the technical contribution of Michael d'Mello (Thinking
Machines
Corporation) is acknowledged),
and on an IBM workstation cluster at SCRI, Florida State
University, Tallahassee, Florida.

\figure{ Energy of a hole
in the ${\rm t-J}$ model, $\epsilon ( {\bf
k})$, vs momentum obtained with the GFMC method on a $12 \times 12$
lattice and ${\rm J/t = 0.4}$ (in units of t). The numerical results are the
full
squares, and the solid line is to guide the eye. Note the flat region
near the X point. The dashed results correspond to numerical results
for the same lattice and ${\rm J/t}$, but including a n.n.n. hopping
term in the Hamiltonian, ${\rm t' = -0.15t}$. The error bars are not
shown but typically they are ${\rm \approx 0.02t}$ at all momenta, with the
exception of the the $\Gamma$ and $M$ points where they are ${\rm \approx
0.20t}$.
\label{fig1}}

\figure{
(a) Energy vs momentum of
the experimentally observed quasiparticle band
obtained using ARPES techniques (from Ref.\cite{dessau}) (open
circles). The dashed line is an extrapolation discussed
in \cite{dessau}. The theoretical
predictions of the present paper
are shown as full squares joined by a solid line. They correspond to
${\rm J/t = 0.4}$, on a $12 \times 12$ cluster.
(b) Density of states obtained from Eq.(2)
(${\rm t-J}$ model, ${\rm J/t=0.4}$) showing the van-Hove singularity.
The unit of energy is ${\rm t}$.
\label{fig2}}

\figure{(a) Occupation number, ${\rm n_k }$, versus momenta at
hole concentration ${\rm x = 0.10}$, ${\rm J/t=0.4}$ and several
equidistant temperatures ranging from
$T=0$ to $T = 0.10 t$. The arrows indicate the direction in which T is growing.
The rapid reduction in the hole population
near ${\bar M}$ is apparent; (b) Fermi surface as defined in the text
corresponding to ${\rm J/t=0.4}$, ${\rm x=0.1}$, and zero temperature;
(c) Same as (b)
but at ${\rm T \approx 0.08t \approx 400K}$. The full squares denote the
position of the saddle-points;
(d) Sign of the Hall coefficient, ${\rm R_H}$, in the plane
temperature - hole density ${\rm (T,x)}$. The solid line separates
hole-like (${\rm R_H} > 0$) from electron-like (${\rm R_H} < 0$) behavior,
at ${\rm J/t = 0.4}$.  The dashed line separates the regime where the
Fermi surface is small and hole-like from the region where the FS is
large and electron-like.
\label{fig3}}

\figure{ (a) Hall coefficient (${\rm R_H}$) (in units of ${\rm 10^{-3}
cm^3 / C}$)
as a function of hole density
(x). The solid line denotes results for the pure ${\rm t-J}$ model
at ${\rm J/t = 0.4}$ and T=0.
The full
squares are experimental results for ${\rm La_{2-x} Sr_x Cu O_4}$ at
${\rm T=50K}$,
taken from Hwang et al. (Ref.\cite{hwang}) ;
(b) ${\rm R_H}$ against temperature at a fixed hole density. The upper
solid curve is the theoretical result obtained at ${\rm J/t = 0.4}$, and
${\rm x \sim 0.12}$. The nearby full squares are experimental results from
Hwang et al. (Ref.\cite{hwang}) corresponding to ${\rm x = 0.125}$. The lower
solid curve and accompanying full squares correspond to theoretical and
experimental results, respectively, obtained at hole density ${\rm x = 0.22}$.
\label{fig4}}



\begin{references}
\bibitem{comm77} In the approach of Trugman\cite{trugman} the change of sign
increasing hole doping was observed at ${\rm x \sim 0.26}$. Since in the
present approach, an accurate dispersion relation is used, then
we expect our results to be more representative of the t-J model.
\end{references}
\end{document}